# Rapid Scout Microscopy of Unstained Tissue Samples in Pathology Protocols with an X-Ray Tomosynthesis Microscope

**Authors**: Han Wen[1,*], David T. Nguyen[1], Thomas C. Larsen[2], Muyang Wang[1], Russel H. Knutsen[1], Zhihong Yang[1], Eric E. Bennett[1], Dumitru Mazilu[1], Zu-Xi Yu[1], Xi Tao[3], Danielle R. Donahue[4], Ahmed Gharib[5], Christopher K. E. Bleck[1], Joel Moss[1], Beth A. Kozel[1], Alan T. Remaley[1]

1. National Heart, Lung and Blood Institute, National Institutes of Health, Bethesda, Maryland, USA. 2. College of Medicine, University of Arizona, Tucson, Arizona, USA. 3. Southern Medical University, Guangzhou, Guangdong, China. 4. Mouse Imaging Facility, National Institutes of Health, Bethesda, Maryland, USA. 5. National Institute of Diabetes and Digestive and Kidney Diseases, National Institutes of Health, Bethesda, Maryland, USA.

**Source of support:** This project is funded by the Intramural Research Program, National Institutes of Health, Division of Intramural Research, National Heart, Lung and Blood Institute, USA.

**Address correspondence to:**

Han Wen
Building 10, Room B1D416
10 Center Drive, MSC 1061
Bethesda, MD 20892, USA
Email: wenh@nhlbi.nih.gov

**Disclosures:** none declared.

**Abstract:** In pathology protocols, each tissue block can generate a large number of sections making it impractical to analyze every section. X-ray microscopy that provides scouting of intact tissue blocks can help pinpoint the relevant structures in 3D space for subsequent analysis, and thus reduce workload and enable further automation downstream. Unlike independent virtual histology studies by laboratory micro computed tomography (CT), routine scout imaging is constrained by a time window of minutes and minimal sample handling to avoid interfering with the histopathology protocols. We found a form of x-ray tomosynthesis used in security screening to be particularly suitable to meet this need when compared to micro CT. When compared to two commercial micro CT scanners, it shortened the scan time from 3 hours to 15 minutes and increased the contrast-to-noise ratio from 3.3 to 14.6. We report the results from various types of human and animal tissue samples, where it served as an integral step of pathology protocols. In an HIV case it provided a new discovery of isolated focal calcification in the internal elastic lamina as the onset of medial calcific sclerosis. It also provided quantitative measurements of calcification in intact samples, which were difficult to obtain by standard pathology procedures. The prospect of continuous and automated screening of many samples in an assembly-line approach is discussed.

## Introduction

The aim of this work is not x-ray based virtual histology, but to meet a frequent need in pathology protocols in both laboratory and clinical settings. The need becomes apparent when we appreciate that a single sample can yield hundreds of histology sections, and that some research studies involve many samples from large cohorts of mice. Staining and analyzing just a fraction of these sections becomes a laborious task. Similarly, the clinical setting forces a trade-off between patient throughput and sampling density in biopsy specimen[1–3]. If a microscopy scan can be performed rapidly and with minimal sample handling, the information helps to focus the pathology analysis and avoid skipping relevant features in a blind search. The need comes with very specific constraints: a time window of 15 minutes to scout the entire sample with minimal sample handling so as not to interfere with the subsequent histology procedure.

X-ray techniques offer the necessary penetration for thick samples. Traditionally, micro CT studies of tissue samples provide high-resolution structural information with the help of x-ray absorbing contrast agents (see references[4–6] for a review of the vast literature). Since the required photon fluence scales up cubically with image resolution, virtual histology by micro CT progresses towards higher resolution at increasing number of hours of scanning and strong contrast staining. Coming from an earlier experience of screening resin blocks for electron microscopy[7], micro CT was also our initial choice for the present application. However, our trials eventually found it incompatible with the present need in two aspects. The first was the scan time being several hours or longer without x-ray contrast staining[6], the second was the issues associated with x-ray contrast staining, including alteration to the sample and the long time required to stain intact samples.

Therefore, there is a need for an alternative method that is specifically designed for rapid scouting of unstained tissue samples. A particular mode of x-ray tomosynthesis that has been deployed for luggage screening at airports was found to meet the need. Tomosynthesis, or sometimes called laminography in the industrial fields, is the process of viewing a scene from a range of angles in order to obtain depth information[8], an enhanced version of stereo vision[9]. In this version of tomosynthesis, the samples are scanned linearly through a stationary imaging system, such as luggages moving through an inspection station on a conveyor belt. This method is widely used in security screening and industrial inspection[10–12] (Adams JA: Automatic warp compensation for laminographic circuit board inspection. US Patent 5687209, 1997) since its proposal in the mid-1990s[13,14].

A key factor that affects efficiency for this particular application is that the standard tissue-embedding cassettes and the small dishes that hold samples in solution all have flat shapes. It means that in a straight-line tomosynthesis scan, the sample can be laid flat under the x-ray focal spot at a close distance (see Fig. 1). In a CT scan, more space between the sample and the x-ray source is needed to allow un-hindered rotation of the sample. A minimal distance between the sample and the x-ray source maximizes the photon flux through the sample. This is the main reason for the improved efficiency of straight-line tomosynthesis for this particular application, when hardware performance such as source power and detector sensitivity being equal. To our knowledge, the present work is the first adaption of the method to tissue microscopy. The resulting microscope design resembles a vertical light microscope (Fig. 1), with an open sample stage to accept a variety of samples. The sample stage is scanned horizontally, while the rest of the microscope are stationary.

The depth of imaging is limited by the penetration of the x-rays through the sample, and therefore hard x-rays are required for typical tissue blocks. The range of depth at which the lateral and depth resolutions are maintained is limited only by the space between the focal spot and the detector, provided that the travel range of the scan covers the span of the x-ray cone beam.

The microscope has been used in various pathology procedures for both human and animal tissue samples. To-date it has scanned 80 tissue samples of various types include 21 human tissue samples. The lateral sizes are up to several centimeters and thickness up to 15 mm. Examples for six different types of samples are reported below. The results from micro CT test on two different commercial scanner models are also described for a comparison of the two approaches for this particular application.

## Materials and Methods

Research on human tissue samples was performed under protocols approved by the Institutional Review Board of the National Heart, Lung and Blood Institute, NIH, USA. Research on animal tissue samples was performed under protocols approved by the institutional Animal Care and Use Committee.

**Samples and study procedures**

To represent a variety of situations, the samples include left anterior descending coronary artery specimens from an HIV patient donor, lung biopsy tissue specimen from a patient with lymphangioleiomyomatosis (LAM) who received a lung transplant, a brain specimen from a mouse haploinsufficient for elastin (Eln+/-), and aorta and heart specimens from ApoE knock-out (ApoE-/-) mice. The study procedure for each sample is described below.

*Human coronary artery sample*: The histopathology protocol of the coronary artery specimens from the HIV patient donor aimed to study the pattern of coronary artery disease in HIV[15]. The specimens were fixed in 10% buffered formalin and processed with Leica ASP-300 tissue processor. They were embedded in paraffin in a tissue-embedding cassette (Fig. 2A). The sample was scanned on a commercial micro CT scanner (Bruker Skyscan 1172) and on the present x-ray microscope for comparison of image quality and speed. Soft tissue visibility was measured by the contrast-to-noise ratio (CNR) between the wall of the blood vessel and the surrounding wax medium: CNR = (signal difference between the vessel wall and the wax medium)/(standard deviation of signal in the wax medium). Locations of interests were identified in 3D space from the x-ray image stack. The block was then sectioned to those locations with a microtome. Slices of 5 to 10 μm thickness were mounted in positively charged slides. The slides were stained with either hematoxylin and eosin (H&E) for soft tissue structure, or von Kossa stain to highlight calcium phosphate deposits while also delineate the structural layers of the wall. The slides were scanned in a Hamamatsu digital slide scanner (NanoZoomer Res 2.0, Shizuoka Pref., Japan).

*Human lung tissue biopsy specimen*: Histopathology of the lung tissue biopsy sample from the LAM patient was focused on the structure of the cyst wall. Patients with LAM develop diffused cysts with thin walls in their lungs. This is accompanied with decreased pulmonary function and diffusion[16]. The lung tissue was fixed in 10% buffered formalin, processed and paraffin embedded in a cassette as described above (Fig. 3A). The tissue block was scanned with the x-ray microscope to identify locations of the cysts and interesting features in their walls. The block was then sectioned to those locations to produce slices of 5 μm thickness. The slices were H&E stained and scanned in the digital slide scanner to obtain color micrographs. The cell types present in the cyst wall were recognized by their appearance with H&E staining.

*Mouse heart sample*: The mouse heart sample was from a one-year old ApoE-/- mouse. The goal of the study was to image the distribution of calcification in the aortic root, and in particular, to detect possible calcification deposits on the aortic valve. The mouse heart was fixed and embedded in paraffin as described above. X-ray microscopy was performed on the tissue block. Calcification deposits were detected by their strong x-ray absorption compared to the surrounding soft tissue[15]. Quantitative mass distribution of calcification was measured with a modified single-energy x-ray absorptiometry method[17].

*Mouse brain sample*: The study of the brain specimen from the Eln+/- mouse focused on exploring possible vascular abnormalities in the brain arising from the genetic modification[18]. The specimen underwent the standard fixation and paraffin-embedding process as described above. The x-ray microscopy scan provided the z-stack of images, by which a custom software was used to create cross-sectional images along the curved paths of blood vessels (Fig. 5B). The cross-section follows landmarks which are defined by the user in the x-ray images.

*Mouse aorta sample*: The mouse aorta sample was from a one-year old ApoE-/- mouse. The aim of the study was to quantitatively evaluate the amount of atherosclerotic plaques and calcification in the aorta. The sample was a full length aorta extending from the aortic arch to the common iliac arteries. In the standard procedure, the aorta is lightly irrigated with 10% buffered formalin for several seconds to preserve it without rubberizing the vessel wall. It is then cut lengthwise and laid open to expose the lumen, stained with Sudan IV red for lipids, and sealed between two microscopy slides in phosphate-buffered saline (PBS) solution. Light microscopy is then used to detect red-stained patches in the lumen. To image and quantify calcification, x-ray microscopy was inserted twice into the procedure. The first was on the intact aorta sample to avoid potential loss of the calcium deposits in the dissection and staining steps. In the x-ray microscopy images, calcification was detected by their strong x-ray absorption compared to the surrounding soft tissue[15]. Quantitative measurement of the mass distribution of calcification used a modified single-energy x-ray absorptiometry method[17]. The sample was immersed in distilled water in a small weigh boat for the x-ray scan (Fig. 6A).

The second x-ray scan was on the microscopy slide of the dissected aorta (Fig. 7A). It was used to obtain a calcification map which could be correlated with the light microscopy image. The total amount of calcification in the slide was measured and compared with that of the intact sample from the first x-ray scan. Plastic microscopy slides (Fisherbrand disposable microscope slides, Fisher Scientific Inc, Hampton, NH) were used for their low x-ray absorption.

**X-ray microscopy method and scan parameters**

The microscope is illustrated in Fig. 1. The sample stage is motorized in the horizontal xy plane under computer control. It is scanned linearly across the x-ray cone beam. The cone beam spans an angle of 80°. During the scan, the sample is sequentially illuminated by x-rays coming from different directions. The series of projection images contain the necessary information to re-construct a stack of depth-resolved cross-sectional images of the sample, called a z-stack.

The field of view in the direction along the scan is adjustable by the travel range of the scan, which is extended for long samples. The x-ray tube has a focal spot of approximately 5 µm at the operating condition of 30 kV/160 µA. The x-ray area detector has a pixel size of 74.8 µm. Each projection image has 3072 × 3840 pixels. Projection images are acquired continuously at a frame rate of 6.9 frames/second. A total of 6250 projections are acquired. The total scan time is 15 minutes. The scan direction (x or y), speed and range are set by the user. The scan range is between 15 mm and 75 mm.

**X-ray microscopy image reconstruction and system resolution**

A reconstructed z-stack generally contains several hundred to several thousand slices. The voxel size is 5 to 7 µm. The number of slices is set by the user to cover the thickness of the sample. A filtered back-projection algorithm is used for reconstruction[19] with modifications to reduce truncation artifacts.

As a tomosynthesis method, the images contain out-of-plane shadows due to the limited coverage of the view angles. The extent of the shadows and the system resolution are quantified in the line spread function and the modulation transfer function. The 10% contrast resolution is 7.3 µm (68.5 line pairs/mm) in the lateral scan direction, and 22.0 µm (22.8 line pairs/mm) in the z direction. Detailed information on these measurements, including the line-spread function and modulation transfer function, are provided in the Supplemental Figure.

**Imaging procedures on two commercial micro CT scanners**

Two micro CT systems, the Skyscan 1172 model and its latest successor the 1272 model (Bruker, Billerica, MA), were used to scan unstained tissue samples embedded in paraffin in standard tissue embedding cassettes.

On the Skyscan 1172 system, the optimal parameters for a 3 hour scan were identified through trials. The scan parameters were x-ray tube setting of 29 kV/167 µA, camera matrix of 3000 by 2096, sample rotation step of 0.20°, 1019 projections over 203.70° rotation angle, 1.767 sec exposure per shot, average over 5 exposures per angle, and total scan time of 3.0 hours. Reconstructed image voxel size was 30.5×30.5×60.9 µm.

On the Skyscan 1272 system, the optimal parameters for a 3 hour scan were determined by a tuning procedure provided by the manufacturer. The scan parameters were determined to be x-ray tube setting of 70 kV/142 µA, camera matrix of 2452 by 1640 after 2x2 binning, sample rotation step of 0.20°, 939 projection angles over 187.8° rotation angle, 1.7 sec exposure per shot, average over 5 exposures per angle, and total scan time of 2 hours 46 minutes. Reconstructed image voxel size was 10.7×10.7×32.1 µm.

# Results

**Human coronary artery sample**

Figure 2B and 2C illustrates the same cross-sectional slice from x-ray microscopy and from one of the micro CT systems, respectively. The contrast-to-noise ratio between the wall of the blood vessel and the surrounding wax medium was 14.64±2.47 for x-ray microscopy at reconstruction pixel size and slice spacing of 7 µm, 3.31±1.21 for the micro CT scan with the Skyscan 1172 system at reconstruction voxel size of 30.5×30.5×60.9 µm, and 1.77±1.09 for the scan with the Skyscan 1272 system at reconstruction voxel size of 10.7×10.7×32.1 µm. The scan times on the three different devices were 15 minutes, 3 hours and 2 hours 46 minutes, respectively.

One of the coronary artery segments is magnified in Fig. 2D. Here the x-ray microscopy scan provided a new finding. Hyper intense dots were found between the intima and media layers of the vessel wall. They indicate micro-calcification deposits of 10 to 20 µm size (yellow arrows in Fig. 2D). As detailed in reference[15], in one tissue block these were found in 12 locations, of which

11 were matched to focal calcification in histology slides (Fig. 2E). In the color micrographs of the slides, the calcium phosphate deposits were stained to a dark brown to black color by the Von Kossa stain. The micrographs showed that the calcification deposits were in the internal elastic lamina (IEL) layer (Fig. 2E). Their sizes ranged from one to several cells. Since they were observed in the intact sample by x-ray microscopy, they were native tissue structures as opposed to particle contaminants that may be introduced in the sectioning and histology procedure. The calcification was identified as the initiation of Monckeberg medial calcific sclerosis, as part of the vascular disease associated with the HIV progression in this patient[15].

**Human lung tissue biopsy specimen**

Figure 3B illustrates a cross-sectional x-ray image of the lung tissue specimen from the LAM patient. It is at the level of 1.76 mm below the surface of the paraffin block. Cysts appear as darker (low density) voids surrounded by brighter rims. Calcified scar tissues appear as highly absorbent of x-rays. One of the cysts is magnified in Fig. 3C. Its wall contains two thickened segments. Color micrograph of the exact location in the H&E stained histology slide is shown in Fig. 3D. The thickened sections are further magnified and shown in the figure insets. By the morphology of the cells, they were recognized as a nodule consisting of smooth muscle cells, and a sedimentation of erythrocytes. In clinical ultra-high resolution CT scans of LAM patients, thickened segments of the cyst wall are often seen with similar appearance to the x-ray micrographs, although at much reduced resolution[20,21]. Results from this study are used to help interpret the clinical CT observations.

**ApoeE -/- mouse heart sample**

Figure 4A is a cross-sectional image of the paraffin-embedded sample from the x-ray microscope. Visible soft tissue structures include the main chambers of the heart, walls of major blood vessels including the aorta and main coronary arteries, valve leaflets and some myocardial fiber structures. Calcification deposits appear as isolated dots and patches of hyper intensity, reflecting their elevated x-ray absorption relative to the background soft tissue. They are concentrated in the aortic root. The 3D disposition of calcification in the aortic wall and the hinges of the valve leaflets is shown in Fig. 4B. A movie of a 360° rotation of the 3D render is available as supplemental material. Quantitative distribution of calcification is represented in a 2D projection view of the aortic root in color scale, which is superimposed onto the x-ray attenuation map in Fig. 4C.

**Eln+/- mouse brain sample**

A typical cross-sectional image of the paraffin-embedded brain sample is shown in Fig. 5A. An unusual feature of the basilar artery at the base of the brain (Fig. 5B) was seen in a digitally constructed cross-section that bisects the vessel lumen along its length (Fig. 5C). The basilar artery appears to be split at mid length into two parallel vessels in the anterior portion of the artery. The appearance was explained by a transverse cross-sectional image of the vessel lumen at the anterior portion (Fig. 5D). The vessel wall is seen to fold inward to form a "V" shaped lumen. As a result, the longitudinal cross-section cuts across the two arms of the "V", which gives the appearance of bifurcation of the basilar artery.

**ApoE-/- mouse aorta sample**

In the fresh aorta sample, a digitally constructed cross-sectional image that bisects the lumen along its length is shown in Fig. 6B. Fatty tissues inside and outside the vessel lumen appear darker due to their lower x-ray attenuation than water. Calcification is seen as hyper intense areas in the aortic arch, extending into the roots of the ascending branches and the upper part of the descending aorta. The areal density of calcification (calcium hydroxyapatite) is shown in a color scale overlay on the attenuation image (Fig. 6C). The total amount of calcification was measured from the x-ray attenuation to be 251 µg.

After the aorta was dissected open and processed into a microscopy slide, light microscopy showed pink/red staining of the yellow fatty plaques on the surface of the lumen (Fig. 7B). The x-ray microscopy z-stack was averaged to produce the x-ray attenuation image of Fig. 7C. Areas of low x-ray attenuation (dark patches) match the lipid-rich areas in the light microscopy image of Fig. 7B. Residual calcification appears as patches of hyper intense x-ray attenuation in Fig. 7C. These are concentrated in the aortic arch area. An overlay of the calcification and the light microscopy image shows the location of calcification (Fig. 7D). The amount of residual calcification was measured to be 46 µg from x-ray attenuation. This was 18% of the amount in the intact sample in Fig. 6. Therefore, some of the calcification was washed away in the procedure of slide preparation.

# Discussion

Since non-invasive scouting of tissue blocks within the pathology workflow share some common requirements with luggage screening at airports and industrial inspection, it is not surprising that a technology developed for the latter[10–14] was also effective for the pathology application. As far as rapid scouting of unstained tissue blocks and fresh samples, the x-ray tomosynthesis microscope demonstrated an inherent efficiency and versatility. It met the need for identifying the depth and location of relevant structures in tissue blocks to guide subsequent histological sectioning and analysis, in a time window of 15 minutes with minimal sample handing. In a particular human tissue study it also made a discovery of the initiation of the Monckeberg medial calcific sclerosis in the internal elastic lamina of the human coronary arteries[15].

In terms of where to insert the scouting step in the pathology workflow, we found that the image contrast of soft tissue structures is higher in the paraffin embedded condition than in the fresh state. It is recommended that scout imaging be performed after paraffin embedding. In cases of vascular disease when an accurate measurement of the mass of calcification or other mineral deposits is required (calcium scoring), x-ray microscopy of fresh samples may be necessary, before the partial loss of mineral particles in the tissue processor[22–25].

In terms of the limitation of this method, it has the inherent trade-off of tomosynthesis. Tomosynthesis is not able to resolve flat horizontal layers unless they have well-defined edges, since the layers overlap completely unless directly viewed from the side. In tomosynthesis imaging, objects above and below a plane cast blurry shadows into the plane. This effect can be seen in the tissue blocks in Fig. 2 - 4, where the plastic grid of the embedding cassette casts shadows into the images. Certain brands of anti-static weigh boats contain numerous particles of additives in their material. The shadows of the particles appear as a wavy background that can be seen in Fig.6B. Nevertheless, experimental data presented here show that the approach has sufficient resolution in the depth direction to meet the particular need of rapid scouting of tissue blocks, namely to identify the depths and locations of relevant structures in the blocks for subsequent histological sectioning and analysis.

With the advent of digital breast tomosynthesis, advanced iterative image reconstruction algorithms have proven effective in suppressing image artifacts[26,27]. Currently computation time is a technical limitation. For example, starting with the 125 GB of raw data of each microscopy scan, it takes 10 to 20 minutes to reconstruct a z-stack of 1000 slices with a basic filtered back-projection algorithm. On the other hand, image reconstruction by nature is amenable to massively-parallel computing. There is room to substantially shorten the computation time by developing GPU-based parallel-processing software. Implementing advanced algorithms to reduce image artifacts while keeping the computation time practical for routine application is a focus of ongoing development work.

Centralized pathology centers often process a large number of samples from multiple sources on a daily basis. A future prospect of the current technology is to serve as a key element in enhanced automation at large centers. Specifically, the possibility of continuous imaging of a stream of standardized samples, such as paraffin-embedded tissue blocks, which passes through the x-ray microscope in indexed carriers on a conveyor belt. A familiar example of this mode of operation is the luggage screening systems at airports. The main challenge will be the rate of image data that flow from the system in continuous operation. The same challenge in luggage screening has spurred the development and deployment of automated threat detection based on artificial intelligence, such as neural network algorithms and machine learning (US Department of Homeland Security: Advanced Integrated Passenger and Baggage Screening Technologies. 2018. https://www.dhs.gov/sites/default/files/publications/TSA%20-%20Advanced%20Integrated%20 Passenger%20Screening%20Technologies.pdf)[28]. The same solution can be envisioned for the pathology application. It is conceivable that machine learning-based software will identify in real time and in 3D space targets for sectioning and staining in each of the stream of tissue blocks that pass through the system, which are relayed to automated sectioning[29] and staining systems that produce the slides for digital slide scanners. The goal of such automation is to reduce workload and enhance accuracy in pathology studies[30].

The contrast of soft tissue structures in x-ray images is generally much lower and more subtle than color micrographs of stained slides. X-ray contrast often involves texture features at the pixel level. Such x-ray based biomarkers have been established in various branches of clinical radiology, but they are yet to be learned in x-ray images of paraffin-embedded tissue blocks, particularly for machine-learning based automation. Newer photon counting and energy resolved x-ray detectors may enhance soft tissue contrast. Ultimately, an "image and learn" approach similar to radiology will be needed through correlative studies between x-ray microscopy and pathology tests to establish the necessary knowledge base.

## Figure Legends

**Figure 1** The x-ray microscope without the radiation enclosure. The size is 38 cm by 30 cm by 60 cm (height). The moving part of the system is the sample stage. It is motorized in the horizontal plane. The sample is scanned across the cone beam in the x or y direction. The cone beam projection provides up to 14x geometric magnification.

**Figure 2** Comparison between a 3-hour scan on a commercial micro CT system and a 15-minute scan on the x-ray tomosythesis microscope, in unstained coronary artery samples from an HIV patient donor. **A**. A photo of the paraffin-embedded artery segments. **B**. A cross-sectional image from the x-ray microscope at 2.85 mm below the surface of the tissue block. The image voxel size

is 7×7×7 µm. Vessel walls appear bright against the wax medium. The background vertical stripes are out-of-plane shadows of the embedding cassette. **C**. Image of the same cross-section from a micro CT scan. The image voxel size is 30.5×30.5×60.9 µm. **D**. A magnified view of the artery segment outlined in panel C. Isolated hyper-dense dots between the intimal and medial layers are indicated by yellow arrows. **E**. Color micrograph of the same location from a histology section with Von Kossa staining. By matching the 3D coordinates with the x-ray image, the hyper-dense dots in the x-ray image are identified as micro calcification in the internal elastic lamina.

**Figure 3** Rapid x-ray tomosynthesis microscopy and histology of a lung tissue biopsy specimen from a patient with the cystic lung disease lymphangioleiomyomatosis (LAM). **A**. A photo of the paraffin-embedded specimen. **B**. An x-ray cross-sectional image at 1.76 mm below the surface of the tissue block. A LAM cyst is highlighted by the dotted rectangle. Blue arrows indicate calcified tissue which appears hyper intense. **C**. In a magnified view of the cyst, two sections of the cyst wall appear thickened, indicated by the red markers. **D**. Histological image of the same location with hematoxylin and eosin staining shows that the thickened sections of the cyst wall are a nodule consisting of smooth muscle cells, and a sedimentation of erythrocytes. The insets are magnified views of the interior of the thickened sections.

**Figure 4** Rapid x-ray tomosynthesis microscopy of an unstained heart specimen from an ApoE -/- mouse. **A**. Cross-sectional image bisecting the left ventricle (LV) and the right ventricle (RV) shows the aortic valve leaflets (red arrows), and hyper-intense calcification deposits in the aortic wall and the hinges of the valve leaflets (blue arrows). **B**. A 3D translucent render of a cutaway section of the aortic root. It shows in red color the distribution of calcification in tissue. A movie of a 360° rotation of the 3D render is available as supplemental material. **C**. In a planar projection view of the aortic root, the areal density of calcification is represented by the color scale.

**Figure 5** Rapid x-ray tomosynthesis microscopy of a paraffin-embedded brain sample from a Eln +/- mouse. **A**. A typical cross-sectional image of the unstained sample. Brightness corresponds to x-ray attenuation. The dark dots and lines in the brain tissue are lumens of blood vessels. **B**. A sketch showing a perspective view of the location of the basilar artery on the underside of the mouse brain. The basilar artery branches into the two cerebral arteries in the posterior, and curves upward into the mid brain at the front. **C**. Digitally constructed cross-sectional image following the course of the basilar artery (the blue surface marked "B" in the sketch in panel B). Vessel walls appear as bright lines due to their higher density than surrounding tissue. The basilar artery appears to split at mid length into two parallel vessels in the anterior portion of the artery (red arrows). The cerebral arteries are indicated by blue arrows. **D**. A transverse cross-sectional image of the anterior portion of the basilar artery along plane "D" in the sketch in panel B. It shows that the vessel wall has folded inward to form a "V" shaped lumen. The vessel wall is outlined by the dotted blue line. As a result, the longitudinal cross section (dotted red line) cuts across the two arms of the "V", which gives the appearance of a split basilar artery in the image in panel C.

**Figure 6** Rapid x-ray tomosynthesis microscopy of an intact mouse aorta sample immersed in water. **A**. A photo of the aorta sample in distilled water in a weigh boat, on the microscope sample

stage. The aortic arch is marked by "+". The common iliac branches at the other end of the aorta can also be seen. **B**. A digitally constructed cross-sectional image along the curved length of the aorta, which bisects its lumen. The inner and outer surfaces of the vessel wall are seen as thin dark lines. Dark patches inside and outside the lumen are fatty tissue with low x-ray attenuation. The aortic arch and surrounding vessels are highly calcified and appear hyper bright (dotted rectangle). The dark shadows surrounding the calcification are due to signal saturation. The wavy background is the shadow of impurities in the material of the weigh boat. **C**. A magnified view of the aortic arch with an overlay of the areal density of calcification in a color scale. The total mass of calcification was 251 µg.

**Figure 7** Rapid x-ray tomosynthesis microscopy of the same aorta sample in Fig. 6 after it was processed into a slide with sudan IV red stain. **A**. Photo of the sample positioned on the x-ray microscope stage. The aorta sample is sealed in a neutral buffer solution between two plastic microscopy slides. **B**. Light microscopy image of the slide shows pink/red staining of the fatty plaques on the surface of the lumen. **C**. X-ray attenuation image shows areas of low attenuation (darker patches indicated by red arrows) that match the lipid-rich areas in B. Residual calcification in the aortic arch appear hyper intense (cyan arrows), which are visible only in the x-ray images. **D**. An overlay of the calcification map and the light microscopy image shows the location of calcifications (cyan colored patches), which are within the lipid-rich areas of the aortic arch. The total mass of calcification was 46 µg.

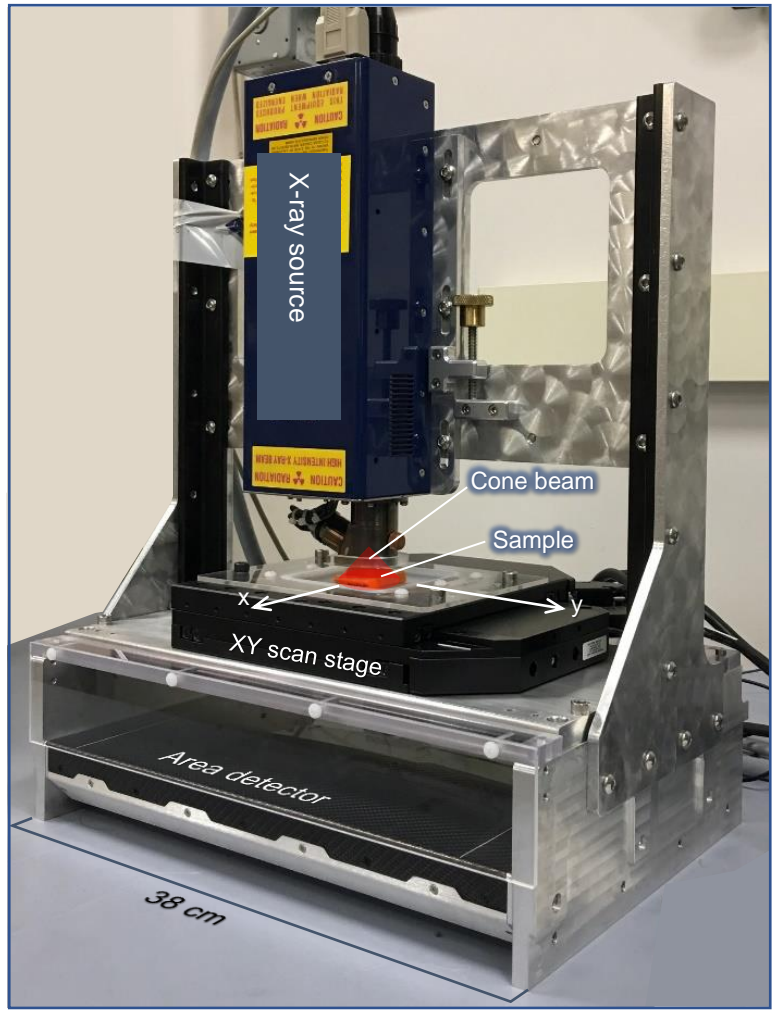

**Figure 1** The x-ray microscope without the radiation enclosure. The size is 38 cm by 30 cm by 60 cm (height). The moving part of the system is the sample stage. It is motorized in the horizontal plane. The sample is scanned across the cone beam in the x or y direction. The cone beam projection provides up to 14x geometric magnification.

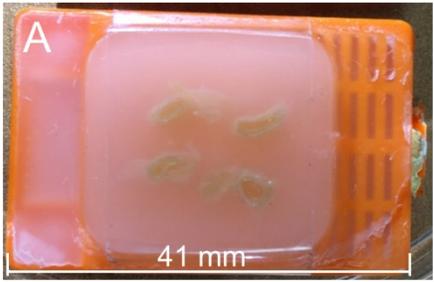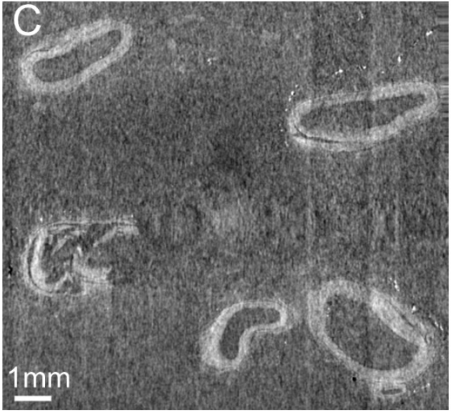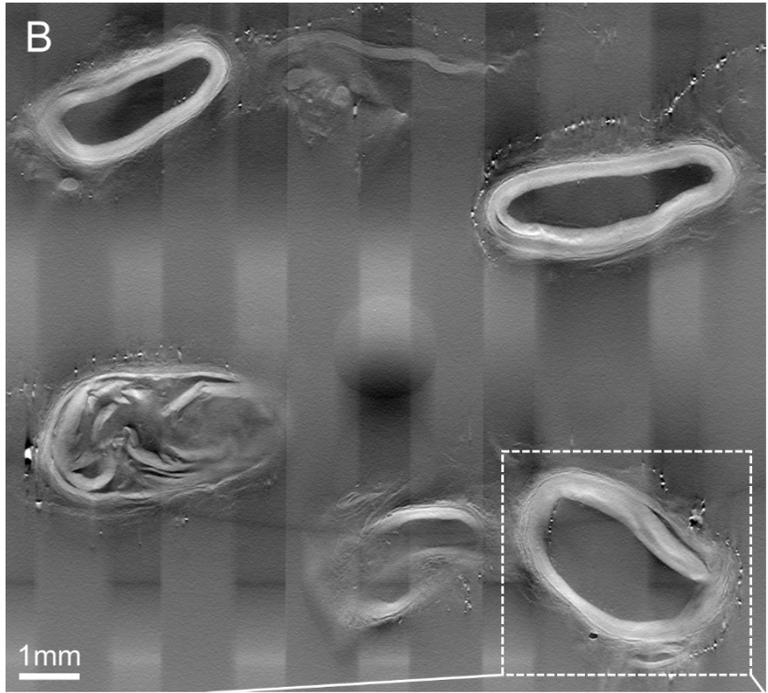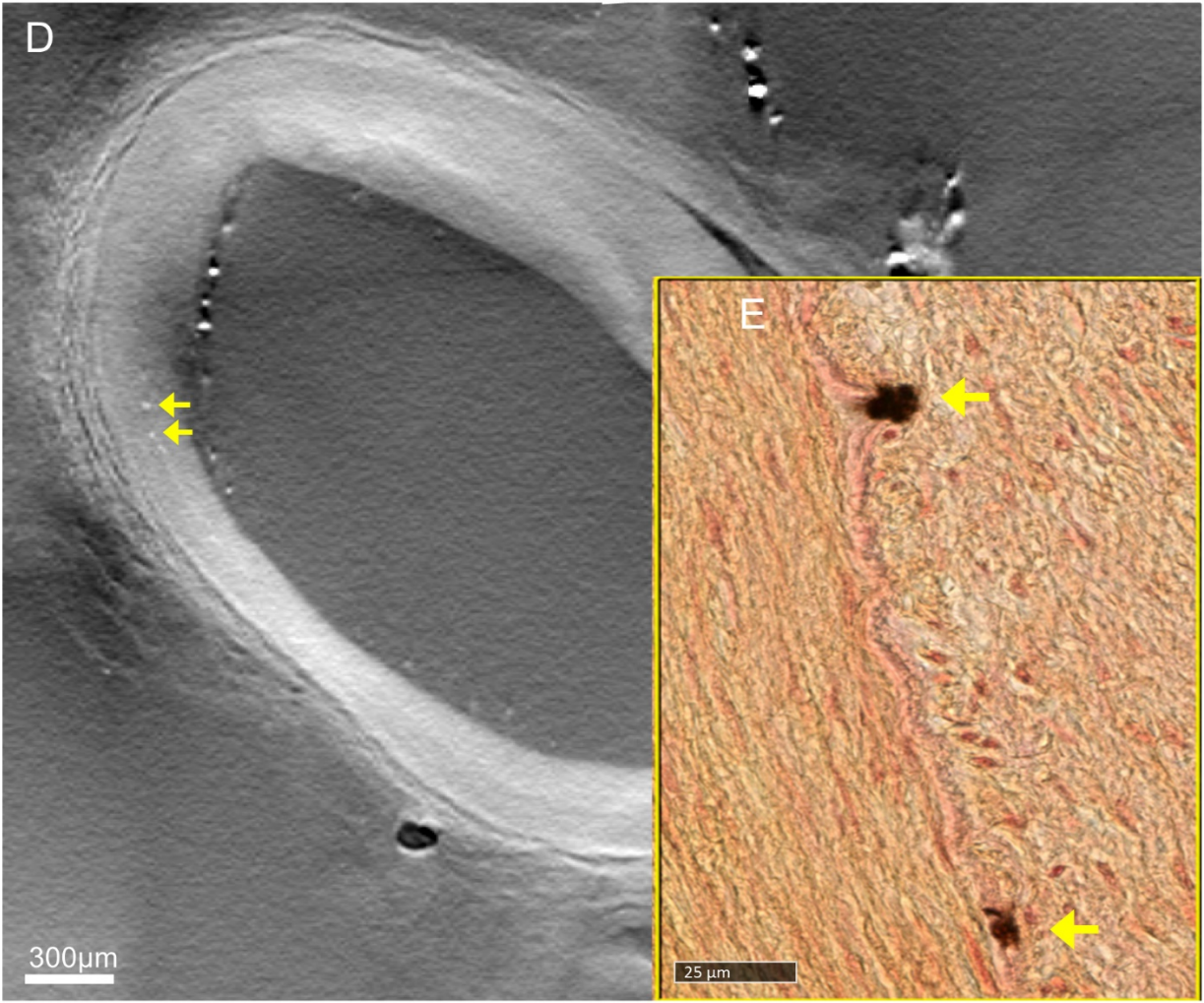

**Figure 2** Comparison between a 3-hour scan on a commercial micro CT system and a 15-minute scan on the x-ray tomosynthesis microscope, in unstained coronary artery samples from an HIV patient donor. **A**. A photo of the paraffin-embedded artery segments. **B**. A cross-sectional image from the x-ray microscope at 2.85 mm below the surface of the tissue block. The image pixel size is 7 μm. Vessel walls appear bright again the wax medium. The background vertical stripes are out-of-plane shadows of the embedding cassette. **C**. Image of the same cross-section from the micro CT scan. The image pixel size is 30.5 μm. **D**. A magnified view of the artery segment outlined in panel C. Isolated hyper-dense dots between the intimal and medial layers are indicated by yellow arrows. **E**. Color micrograph of the same location from a histology section with Von Kossa staining. By matching the 3D coordinates with the x-ray image, the hyper-dense dots in the x-ray image are identified as micro calcification in the internal elastic lamina.

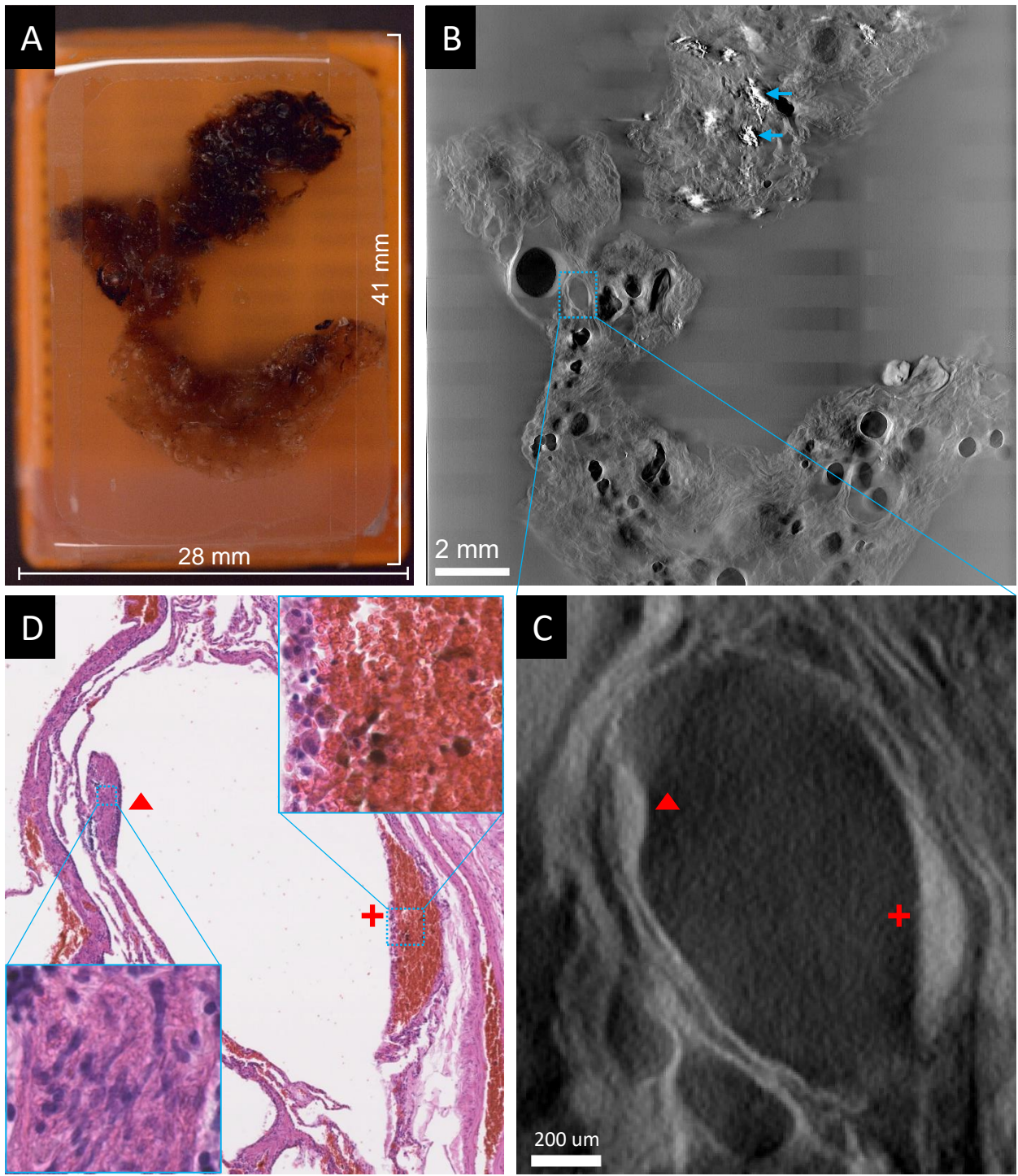

**Figure 3** Rapid x-ray tomosynthesis microscopy and histology of a lung tissue biopsy specimen from a patient with the cystic lung disease lymphangioleiomyomatosis (LAM). **A**. A photo of the paraffin-embedded specimen. **B**. An x-ray cross-sectional image at 1.76 mm below the surface of the tissue block. A LAM cyst is highlighted by the dotted rectangle. Blue arrows indicate calcified tissue which appears hyper intense. **C**. In a magnified view of the cyst, two sections of the cyst

wall appear thickened, indicated by the red markers. **D**. Histological image of the same location with hematoxylin and eosin staining shows that the thickened sections of the cyst wall are a nodule consisting of smooth muscle cells, and a sedimentation of erythrocytes. The insets are magnified views of the interior of the thickened sections.

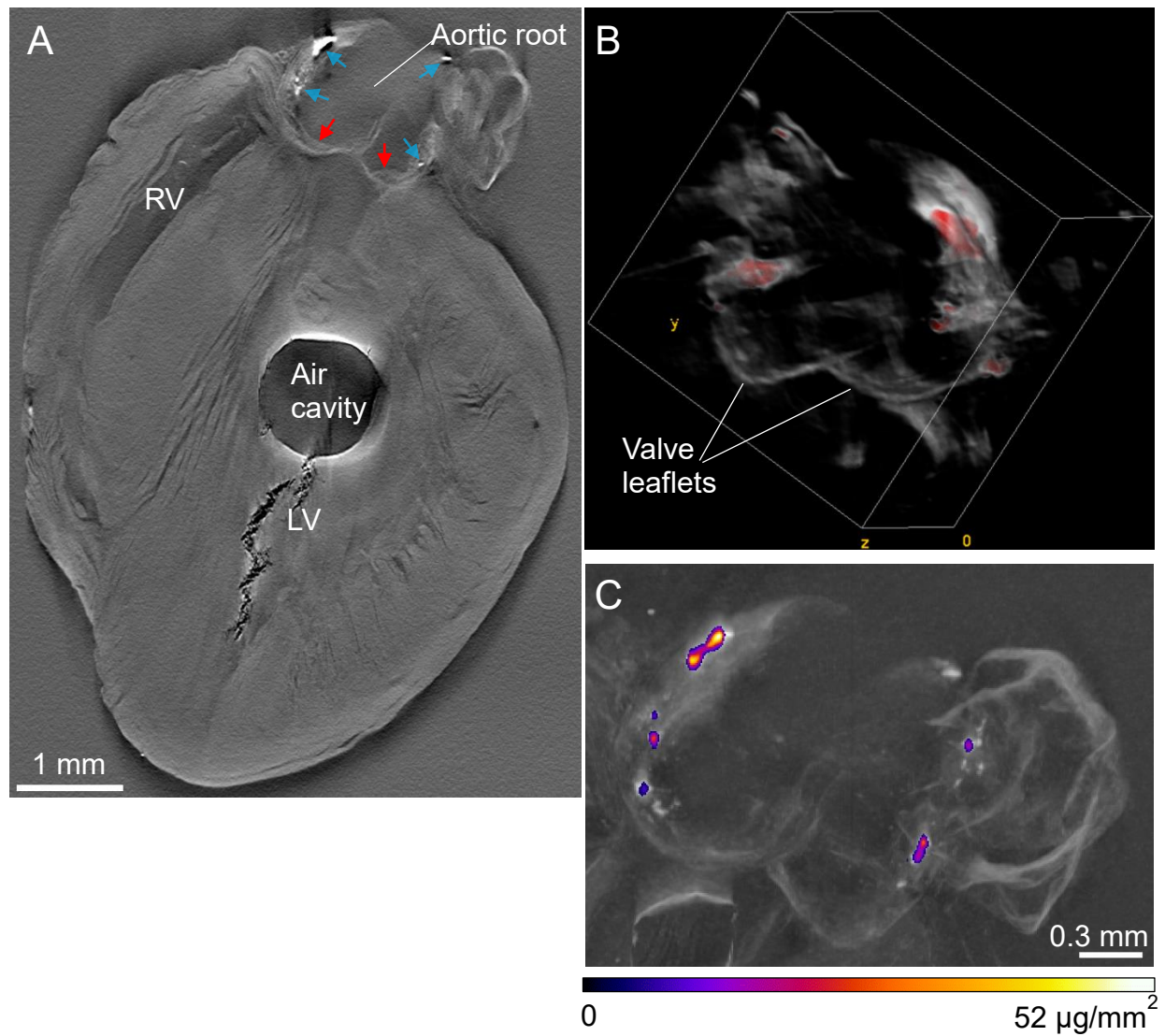

**Figure 4** Rapid x-ray tomosynthesis microscopy of an unstained heart specimen from an ApoE -/- mouse. **A**. Cross-sectional image bisecting the left ventricle (LV) and the right ventricle (RV) shows the aortic valve leaflets (red arrows), and hyper-intense calcification deposits in the aortic wall and the hinges of the valve leaflets (blue arrows). **B**. A 3D translucent render of a cutaway section of the aortic root. It shows in red color the distribution of calcification in tissue. A movie of a 360° rotation of the 3D render is available as supplemental material. **C**. In a planar projection view of the aortic root, the areal density of calcification is represented by the color scale.

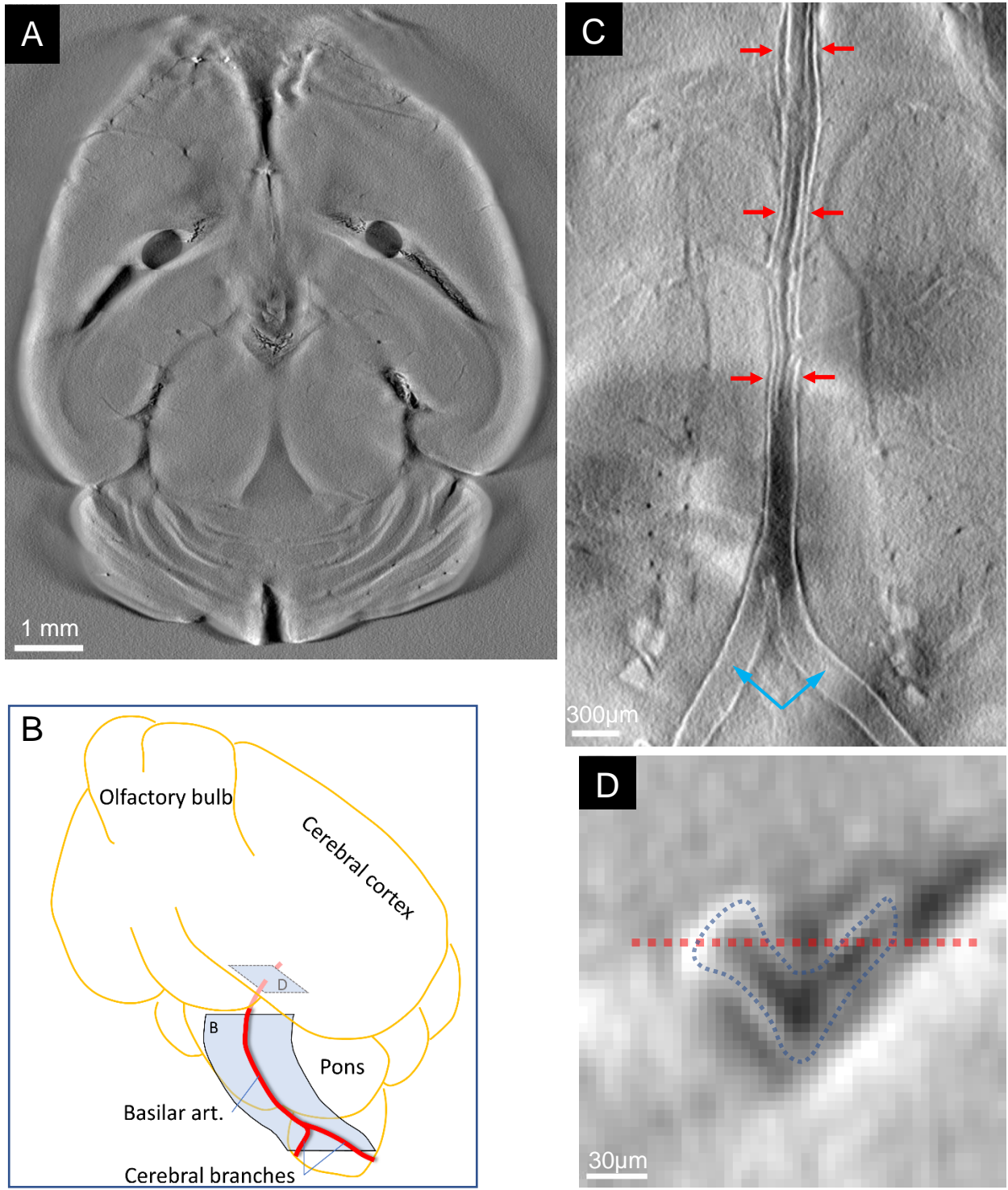

**Figure 5** Rapid x-ray tomosynthesis microscopy of a paraffin-embedded brain sample from a Eln +/- mouse. **A**. A typical cross-sectional image of the unstained sample. Brightness corresponds to x-ray attenuation. The dark dots and lines in the brain tissue are lumens of blood vessels. **B**. A sketch showing a perspective view of the location of the basilar artery on the underside of the mouse brain. The basilar artery branches into the two cerebral arteries in the posterior, and curves upward into the mid brain at the front. **C**. Digitally constructed cross-sectional image following the course of the basilar artery (the blue surface marked "B" in the sketch in panel B). Vessel walls appear as bright lines due to their higher density than surrounding tissue. The basilar artery appears to split at mid length into two parallel vessels in the anterior portion of the artery (red arrows). The cerebral arteries are indicated by blue arrows. **D**. A transverse cross-sectional image of the anterior portion of the basilar artery along plane "D" in the sketch in panel B. It shows that the vessel wall has folded inward to form a "V" shaped lumen. The vessel wall is outlined by the dotted blue line. As a result, the longitudinal cross section (dotted red line) cuts across the two arms of the "V", which gives the appearance of a split basilar artery in the image in panel C.

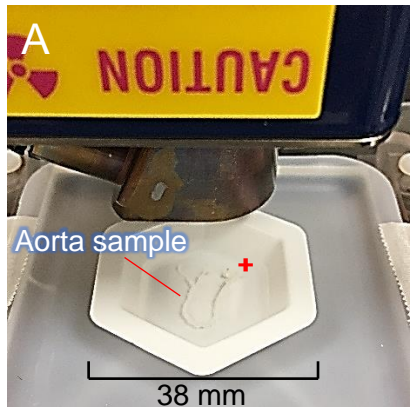
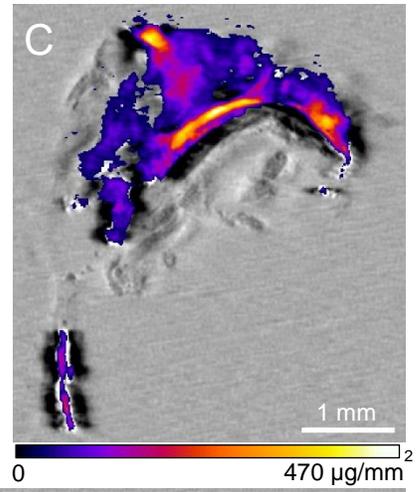
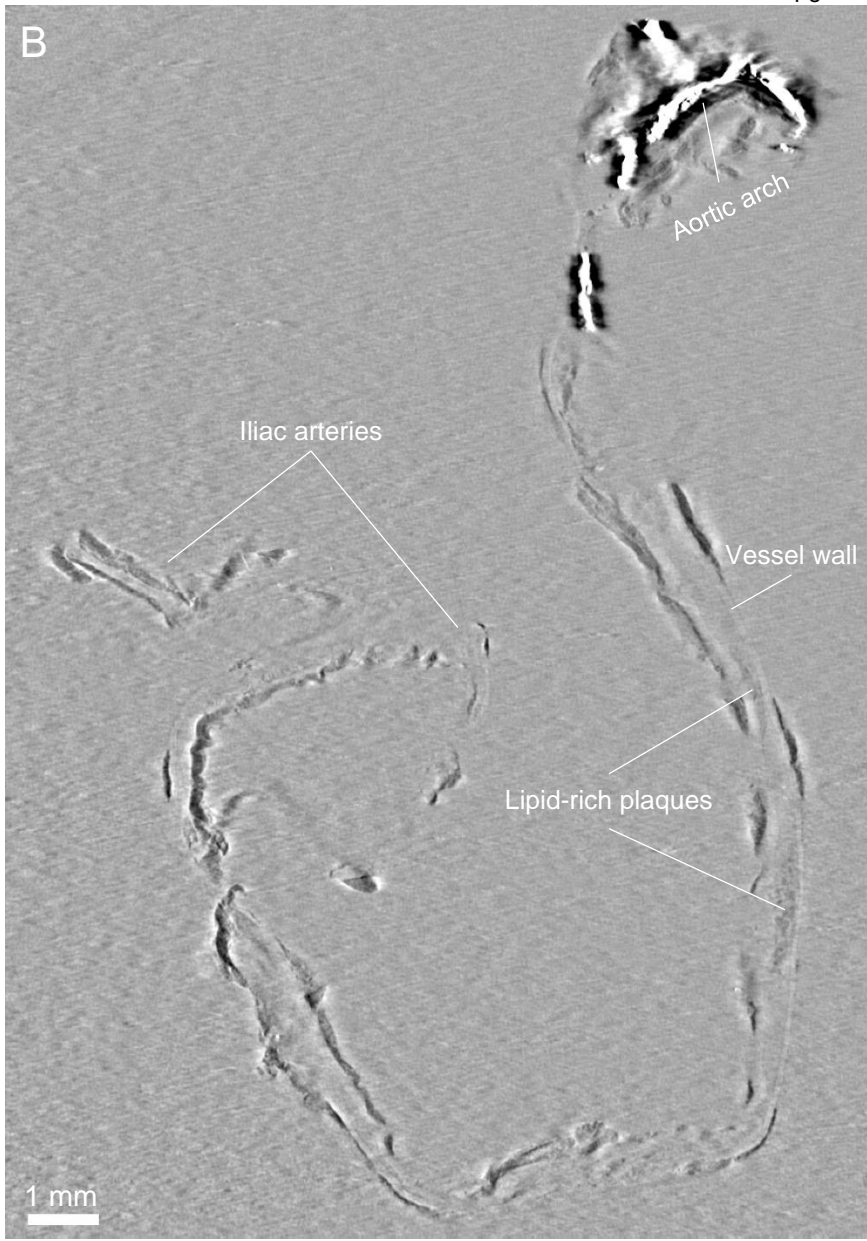

**Figure 6** Rapid x-ray tomosynthesis microscopy of an intact mouse aorta sample immersed in water. **A**. A photo of the aorta sample in distilled water in a weigh boat, on the microscope sample stage. The aortic arch is marked by "+". The common iliac branches at the other end of the aorta can also be seen. **B**. A digitally constructed cross-sectional image along the curved length of the aorta, which bisects its lumen. The inner and outer surfaces of the vessel wall are seen as thin dark lines. Dark patches inside and outside the lumen are fatty tissue with low x-ray attenuation. The aortic arch and surrounding vessels are highly calcified and appear hyper bright. The dark shadows surrounding the calcification are due to signal saturation. The wavy background is the shadow of impurities in the material of the weigh boat. **C**. A magnified view of the aortic arch with an overlay of the areal density of calcification in a color scale. The total mass of calcification was 251 µg.

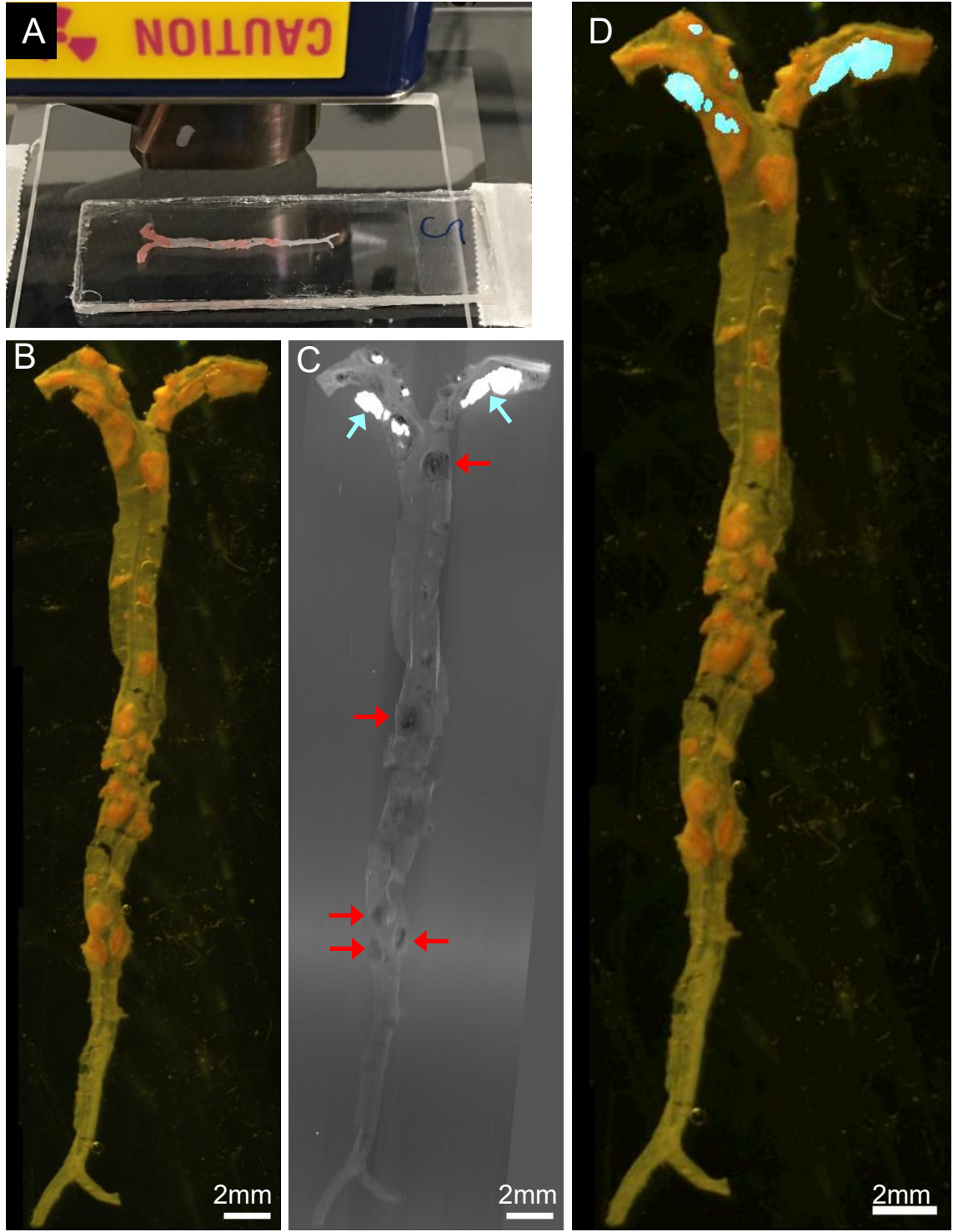

**Figure 7** Rapid x-ray tomosynthesis microscopy of the same aorta sample in Fig. 6 after it was processed into a slide with sudan IV red stain. **A**. Photo of the slide positioned on the x-ray microscope stage. **B**. Light microscopy image of the slide shows pink/red staining of the fatty plaques on the surface of the lumen. **C**. X-ray attenuation image shows areas of low attenuation

(darker patches indicated by red arrows) that match the lipid-rich areas in B. Residual calcification in the aortic arch appear hyper intense (cyan arrows), which are visible only in the x-ray images. **D**. An overlay of the calcification map and the light microscopy image shows the location of calcifications (cyan colored patches), which are within the lipid-rich areas of the aortic arch. The total mass of calcification was 46 µg.

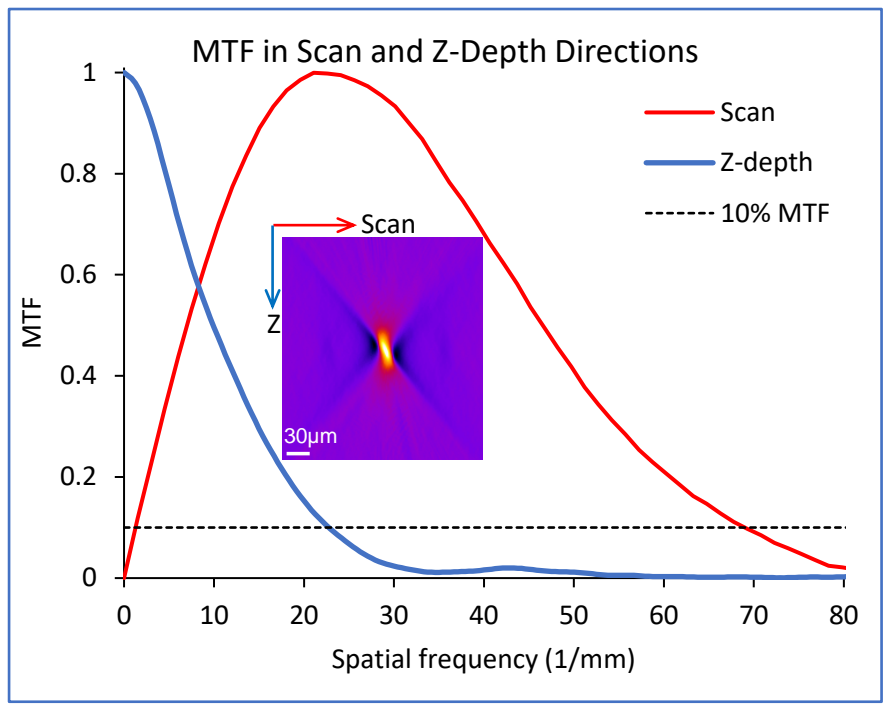

**Supplemental Figure**: Measurement of the line spread and modulation transfer functions of the x-ray optical sectioning microscope. The measurements were from a tungsten wire of 5 µm diameter. Referring to Fig. 1 of the main text, the wire was laid horizontally on the sample stage along the y direction. It was at 9 mm of vertical distance from the x-ray source (focal) spot, which provided 14x magnification. The stage was scanned in the x direction. The z-stack of images were reconstructed at 1.5 µm in-plane pixel size and 1.5 µm slice spacing. The line spread function (LSF) across the wire in the scan(x)-z plane is shown in the inset. Modulation transfer functions (MTF) in the scan and z-depth directions were obtained from Fourier transformation of the x and z cross-section profiles of the LSF, respectively. These are plotted in the figure. The 10% level of relative contrast is marked by the dotted line. The measured 10% contrast resolution in the scan and z-depth directions were 7.3 µm (68.5 line pairs/mm) and 22.0 µm (22.8 line pairs/mm), respectively.